\documentclass[12pt]{article}
\usepackage{amssymb}
\usepackage[T1]{fontenc}
\usepackage[utf8]{inputenc}
\usepackage[intlimits]{amsmath}
\usepackage{enumerate}
\usepackage[dvips]{graphicx}
\usepackage[all]{xy}

\begin{document}
\begin{center}\Huge
Integrable relativistic systems given by Hamiltonians with momentum-spin-orbit
coupling
\end{center}
\begin{center}\large
Alina Dobrogowska  and Anatol Odzijewicz
\end{center}\begin{center}
Institute of Mathematics, University of Białystok\\
Lipowa 41, 15-424 Białystok, Poland
E-mail: alaryzko@alpha.uwb.edu.pl and aodzijew@uwb.edu.pl
\end{center}
\begin{center}
{\bf Abstract}
\end{center}
In the paper we investigate the evolution of the relativistic particle (massive
and massless) with spin defined by Hamiltonian containing the  terms with
momentum-spin-orbit coupling. We integrate the corresponding Hamiltonian
equations in quadratures and express their solutions in terms of elliptic
functions.

\section{Introduction}

Deformation of Lie algebra structure on the fixed vector space $\mathfrak{g}$
is known to provide a construction of a family of Lie--Poisson  brackets on
$C^{\infty}(\mathfrak{g}^*)$, where $\mathfrak{g}^*$ is the vector space dual to
$\mathfrak{g}$. If this family contains a pencil of brackets one obtains
bi--Hamiltonian  structure on $\mathfrak{g}^*$ which leads to the construction
of an integrable hamiltonian hierarchy, e.g. see  \cite{3}, \cite{10}. In the  paper
\cite{7} we investigated such infinite parameter deformation of the Lie--Poisson
structure on the ideal ${\cal L}_+^2({\cal H})$ of the upper triangular
Hilbert--Schmidt operators acting on the real Hilbert space ${\cal H}$, which
has the  properties mentioned above. Applying results obtained in \cite{7} to
the case when $\dim{\cal H}=5$ we find an integrable hamiltonian system, see
equations (\ref{17}-\ref{19}), on the vector space ${\cal L}_+$ of upper
triangular $(5\times 5)$--matrices which one can consider as a dual space to
the Lie algebra ${\cal{E}}_a(1,3)$. This Lie algebra is defined as a one
parameter $a\in [-1,1]$ deformation of Euclidean Lie algebra ($a=1$) in four
dimensions. When $a=-1$ we obtain Poincare algebra and   the Galileo
algebra for $a=0$.

In Section 2 we solve equations (\ref{17}-\ref{19}) in quadratures expressing
their solutions in terms of the first--coordinate $W_0(t)$ of Pauli--Lubansky
four-vector. The dependence of $W_0(t)$ on the evolution parameter $t$ is
also studied.

Using the twistor description of Minkowski space-time (e.g. see \cite{11},
\cite{4}) we investigate  in Section 3 the time-evolution of the position vector
$\vec{X}(t)$ of the massive relativistic particle with a spin, see formula
(\ref{67k}). In Section 3 we also describe  the time dependence of the twistor
coordinates for the massless particle with non--zero helicity, see formula
(\ref{b47b}).

The physical sense of the hamiltonian relativistic system studied in the paper
is hidden in the form of  Hamiltonian (\ref{16}) which gives a coupling between
the momentum,  angular momentum and spin of the relativistic particle.

 In the Euclidean case, i.e. when $a=1$, one has a model with a rather different geometry. Namely,
 from (\ref{12}), (\ref{13}) and (\ref{b4cc}) we find that solution of (\ref{17}-\ref{19}) describes the evolution of a point on the
  bundle $T\mathbb{S}^3$ tangent to the three--dimensional sphere.

\section{Bi--Hamiltonian structure on ${\cal L}_+$ and the related integrable systems}

Let us start our consideration with defining the pencil of metric tensors
\begin{equation}
 \label{b1}
ds_a^2=\eta_{\mu\nu}^adx^{\mu}dx^{\nu}:=a(dx^0)^2+(dx^1)^2+(dx^2)^2+(dx^3)^2,
\end{equation}
where $a\in[-1,1]$, on the four-dimensional affine space $\mathbb{E}_a^{1,3}$
with coordinates $x^{\mu}$, $\mu=0,1,2,3$. We will denote the symmetry group
of $\left(\mathbb{E}_a^{1,3},ds_a^2\right)$ by $E_a(1,3)$ and its Lie algebra by
${\cal{E}}_a(1,3)$, respectively. Using $(5\times 5)$--matrix representation we
find that $g\in E_a(1,3)$ iff
\begin{equation}
 \label{a6}
 g=\left(
 \begin{array}{c|c}
  1 & 0^{\intercal}\\
  \hline
  \tau & \Lambda
 \end{array}\right),
\end{equation}
where $\tau\in\mathbb{R}^4$, $\mathbb{R}^4\ni 0$ is zero vector and
$\Lambda \in Mat_{4\times 4}(\mathbb{C})$ satisfies
\begin{equation}
 \label{a7}
 \Lambda \eta^a \Lambda^{\intercal}=\eta^a.
\end{equation}
Any $\chi\in{\cal{E}}_a(1,3)$ is of the form
\begin{equation}
 \chi=\left(
 \begin{array}{c|c}
  0 & 0^{\intercal}\\
  \hline
  y & X
 \end{array}\right),
\end{equation}
where $y\in\mathbb{R}^4$ and $X \in Mat_{4\times 4}(\mathbb{R})$ satisfies
\begin{equation}
 X\eta^a + \eta^aX^{\intercal}=0.
\end{equation}
The vector space  ${\cal{E}}_a(1,3)^{*}$ dual to the Lie algebra  ${\cal{E}}_a(1,3)$ we will identify with the vector
 space ${\cal L}_+$ of strictly upper triangular $(5\times 5)$--matrices. The pairing between $\rho\in {\cal L}_+$ and
 $\chi\in{\cal{E}}_a(1,3)$ is given  by
\begin{equation}
 \label{b2}
 \left< \chi,\rho\right>:=Tr\left( \chi\rho\right).
\end{equation}
Taking
\begin{equation}
\label{1}
\rho=\left(\begin{array}{ccccc}
0 & P^0 & P^1 & P^2 & P^3\\
0 & 0 & L_1 & L_2 & L_3\\
0 & 0 & 0 & J_3 & -J_2 \\
0 & 0 & 0 & 0 & J_1\\
0 & 0 & 0 & 0 & 0
\end{array}\right)
\end{equation}
and  defining
\begin{equation}
\label{b1a}
M_{0k}=-M_{k0}:=L_k,\;\;\;\;\; M_{kl}:=\epsilon_{kln}J_n,
\end{equation}
where $k,l,n= 1,2,3$, we find that Lie--Poisson bracket $\{\cdot ,\cdot\}_a$ of
$f,g\in C^{\infty}\left({\cal L}_+\right)$ is expressed as follows
\begin{equation}
 \label{a10}
 \{ f,g\}_a=\frac{1}{2}\eta^a_{\varrho\varphi}M_{\mu\nu}
 \left(\dfrac{\partial f}{\partial M_{\varrho\mu}}
 \dfrac{\partial g}{\partial M_{\varphi\nu}}-
 \dfrac{\partial f}{\partial M_{\varphi\nu}}\dfrac{\partial g}{\partial M_{\varrho\mu}}\right)+
\end{equation}
$$
+\eta^a_{\varphi\mu}P^{\mu}
\left(\dfrac{\partial f}{\partial P^{\nu}}
 \dfrac{\partial g}{\partial M_{\varphi\nu}}-
 \dfrac{\partial f}{\partial M_{\varphi\nu}}\dfrac{\partial g}{\partial P^{\nu}}\right)=
 $$
 $$=
aP^0\bigg( \frac{\partial f}{\partial\vec{P}}\cdot
\frac{\partial g}{\partial \vec{L}}
-\frac{\partial f}{\partial \vec{L}}\cdot\frac{\partial g}{\partial \vec{P}}\bigg)+
$$
$$+
\vec{J}\cdot\bigg( a\left(\frac{\partial f}{\partial \vec{L}}\times \frac{\partial g}{\partial \vec{L}}\right)+
\frac{\partial f}{\partial\vec{J}}\times \frac{\partial g}{\partial\vec{J}}\bigg)+
$$$$+
\frac{\partial g}{\partial P^0}\vec{P}\cdot
\frac{\partial f}{\partial \vec{L}}
-\frac{\partial f}{\partial P^0}\vec{P}\cdot \frac{\partial g}{\partial \vec{L}}+
$$$$+
\vec{P}\cdot\bigg(\frac{\partial f}{\partial\vec{P}}\times
\frac{\partial g}{\partial \vec{J}}
+\frac{\partial f}{\partial \vec{J}}\times \frac{\partial g}{\partial \vec{P}}\bigg)+
$$$$+
\vec{L}\cdot\bigg(\frac{\partial f}{\partial\vec{L}}\times
\frac{\partial g}{\partial \vec{J}}
+\frac{\partial f}{\partial \vec{J}}\times \frac{\partial g}{\partial \vec{L}}\bigg).
$$
 Note that one has the following two invariants (Casimir functions) of the coadjoint representation
 \begin{align}
 \label{12}  c_1 & = \eta_{\mu\nu}^aP^{\mu}P^{\nu},\\
\label{13}   c_2 & =\eta_{\mu\nu}^aW^{\mu}W^{\nu},
\end{align}
where
\begin{align}
 &\label{b3} W^0=-\vec{J}\cdot \vec{P},\\
 & \label{b4} \vec{W}=aP^0\vec{J}+\vec{L}\times \vec{P},
\end{align}
while
\begin{equation}
 \label{b4cc}
  \eta_{\mu\nu}^aP^{\mu}W^{\nu}=0.
\end{equation}
The coadjoint representation of $E_a(1,3)$ on the dual space of ${\cal L}_+$ has the form
\begin{equation}
 \label{b5}
 Ad_g^{*}(P,M)=\left( (\eta^a)^{-1} \Lambda \eta^a P,\right.
\end{equation}
$$\left.
 \Lambda\left(\pi_+\left(M\right)-\eta^a\pi_+\left(M^{\top}\right)(\eta^a)^{-1}\right)\Lambda^{-1}+\tau P^{\top}\Lambda^{-1}-\Lambda\eta^aP\tau^{\top}(\eta^a)^{-1}\right),
$$
\begin{equation}
 Ad_g^{*}(W)=(\eta^a)^{-1} \Lambda \eta^a W,
\end{equation}
where  we represent $\rho\in {\cal L}_+\cong {\cal E}_a(1,3)^{*}$ by the  four-momentum $P=(P_{\mu})$ and the angular momentum $M=(M_{\mu\nu})$ defined in (\ref{1}) and (\ref{b1a}).

We note also that for $a,b\in [-1,1]$ the Poisson brackets $\{\cdot, \cdot\}_a$
and $\{\cdot, \cdot\}_b$ define bi--Hamiltonian structure on ${\cal L}_+$, i.e.
their linear combination $\{\cdot, \cdot\}_a+\epsilon \{\cdot, \cdot\}_b$,
$\epsilon\in\mathbb{R}$, is also a Poisson bracket
 on ${\cal L}_+$, see \cite{7}.

 Thus we obtain that Casimirs of $\{\cdot, \cdot\}_b$:
 \begin{align}
 \label{14}  h_1 & =b(P^0)^2+\vec{P}\cdot\vec{P},\\
\label{15}    h_2 & =b\left(\vec{P}\cdot\vec{J}\right)^2+
\left(bP^0\vec{J}+\vec{L}\times \vec{P}\right)^2
\end{align}
 are the integrals of motion being in involution with respect to the Poisson bracket $\{\cdot, \cdot\}_a$.

 Hamiltonian equations associated with the Hamiltonian
 \begin{align}
\label{16}
h= \frac{1}{2}\left(c h_1+dh_2\right)=
\end{align}
$$
=\frac{c}{2}\bigg(b(P^0)^2+\vec{P}\cdot\vec{P}\bigg)+
\frac{d}{2}\bigg(b\left(\vec{P}\cdot\vec{J}\right)^2+
\left(bP^0\vec{J}+\vec{L}\times \vec{P}\right)^2 \bigg)=
$$$$
=\frac{1}{2}(b-a)\left(c(P^0)^2+d(b-a)(P^0)^2\vec{J}^2-\frac{d}{a}\vec{W}^2+2dP^0\vec{W}\cdot\vec{J}\right),
$$
where $c,d\in\mathbb{R}$ are as follows
\begin{align}
\label{17} \frac{dP^0}{dt}= & \{P^0,h\}_a=0,\\
\label{20} \frac{d\vec{J}}{dt}= & \{\vec{J},h\}_a= 0,\\
\label{18} \frac{d\vec{P}}{dt}= & \{\vec{P},h\}_a=(b-a)dP^0\left(\vec{P}\times \left(\vec{P}\times\vec{L}\right)+bP^0\vec{J}\times\vec{P}\right),\\
\label{19}  \frac{d\vec{L}}{dt}=  & \{\vec{L},h\}_a= (b-a)\left(cP^0\vec{P}+bdP^0\vec{J}^2\vec{P}+dP^0\vec{L}\times\left(\vec{P}\times\vec{L}\right)+\right.\\
& \left.+d\vec{P}^2\vec{J}\times\vec{L}-d\left(\vec{P}\cdot\vec{L}\right)\vec{J}\times\vec{P}+bd(P^0)^2\vec{J}\times\vec{L}\right)\nonumber.
\end{align}

In order to solve these equations it suffices to possess four functionally
independent integrals of motion. We choose $P^0$ and $\vec{J}$ as these
integrals. Using the variables $(\vec{P},\vec{W})$ we rewrite (\ref{18}) and
(\ref{19}) in the form
\begin{align}
\label{28} \frac{d\vec{P}}{dt}= & (b-a)dP^0\left(-\vec{P}\times \vec{W}+(b-a)P^0\vec{J}\times\vec{P}\right),\\
\label{29}  \frac{d\vec{W}}{dt}=  & (b-a)d\left(\left(\vec{P}\cdot\vec{J}\right)\vec{P}\times\vec{W}+b(P^0)^2\vec{J}\times\vec{W}+aP^0\left(\vec{P}\cdot\vec{J}\right)\vec{J}\times\vec{P}\right).
\end{align}

Now let us introduce  new variables
\begin{align}
\label{34} & y:=\vec{J}\cdot\vec{W},\\
\label{35} & z:=\vec{J}\cdot\left(\vec{P}\times\vec{W}\right).
\end{align}
From (\ref{28}) and (\ref{29}) we find that these variables and $W^0$ satisfy  the following equations
\begin{align}
\label{36} \frac{dW^0}{dt}= & (b-a)dP^0z,\\
\label{37}  \frac{dy}{dt}=  &- (b-a)dW^0z,\\
\label{38} \frac{dz}{dt}= &
-(b-a)dW^0\left(c_2P^0+c_1aP^0\vec{J}^2-\left(c_1+a(P^0)^2\right)y\right),
\end{align}
which can be integrated in quadratures:
\begin{align}
\label{41}  & t+t_0=\int\dfrac{dW^0}{(b-a)d\sqrt{\frac{-c_1}{4}(W^0)^4+
\frac{c_1(h_2-c_2-(b^2-a^2)(P^0)^2\vec{J}^{\;2})}{2(b-a)}
(W^0)^2+\beta}},\\
\label{40} & y(t)=-\frac{1}{2P^0}(W^0)^2(t)+\frac{h_2-c_2}{2P^0(b-a)}-\frac{b-a}{2}P^0\vec{J}^{\;2},\\
\label{39} & z(t)=\frac{1}{P^0}\sqrt{\frac{-\left(c_1+a(P^0)^2\right)}{4}(W^0)^4(t)+
\frac{c_1(h_2-c_2-(b^2-a^2)(P^0)^2\vec{J}^{\;2})}{2(b-a)}
(W^0)^2(t)+\beta}.
\end{align}
Without loss of generality we can assume $\vec{J}=(0,0,J)$ and obtain
\begin{align}
\label{b6}
& P^3=-\dfrac{1}{J}W^0,\nonumber\\
& W^3=\dfrac{1}{J}\left(-\frac{1}{2P^0}(W^0)^2+\frac{h_2-c_2}{2P^0(b-a)}-\frac{b-a}{2}P^0J^2\right),\\
& (P^1)^2+ (P^2)^2=c_1-a(P^0)^2-\dfrac{1}{J^2}(W^0)^2,\nonumber\\
& (W^1)^2+  (W^2)^2=c_2-a(W^0)^2-\dfrac{1}{J^2}\left(-\frac{1}{2P^0}(W^0)^2+\frac{h_2-c_2}{2P^0(b-a)}-\frac{b-a}{2}P^0J^2\right)^2.\nonumber
\end{align}
After passing to polar coordinates
\begin{align}
\label{b7}
 & P^1=\sqrt{(P^1)^2+ (P^2)^2}\cos \varphi, & & P^2=\sqrt{(P^1)^2+ (P^2)^2}\sin \varphi,\\
  &W^1=\sqrt{(W^1)^2+ (W^2)^2}\cos \psi, & & W^2=\sqrt{(W^1)^2+ (W^2)^2}\sin \psi\nonumber
\end{align}
from (\ref{28}), (\ref{29}) we get
\begin{align}
& \label{b8}
\dfrac{d\varphi}{dt}=(b-a)d_2P^0\left(bP^0J+\frac{y-aP^0J^2}{J}-\frac{(W^0)^2(y+aP^0J^2)}{(W^0)^2-c_1J^2+aJ^2(P^0)^2}\right),\\
&\label{b9}\dfrac{d\psi}{dt}=(b-a)dP^0\left(bP^0J+\frac{y^2-a^2(P^0)^2J^4}{JP^0(c_2J^2-aJ^2(W^0)^2-y^2)}-\frac{1}{P^0J}(W^0)^2\right).
\end{align}
Using formulas  (\ref{b6}), (\ref{b7}), (\ref{b8}) and (\ref{b9}) we find
that the solutions $\vec{W}(t)$, $\vec{P}(t)$ of equations (\ref{28}), (\ref{29})
are expressed by first--coordinate of the spin four--vector  $W^0(t)$ which is
an elliptic function of $t$ defined in (\ref{41}).

Now, we express the solution $\vec{L}=\vec{L}(t)$ by the functions
$\vec{W}(t)$ and $\vec{P}(t)$. To this end we take the vector product of both
sides of (\ref{b4}) and apply the identity
\begin{equation}
 \label{20a} \vec{a}\times\left(\vec{b}\times\vec{c}\right)=\left(\vec{a}\cdot\vec{c}\right)\vec{b}-\left(\vec{a}\cdot\vec{b}\right)\vec{c}.
\end{equation}
This gives
\begin{equation}
 \label{b10}
 \vec{L}=\frac{1}{\vec{J}\cdot\vec{P}}\left(\vec{J}\times\vec{W}+(\vec{J}\cdot\vec{L})\vec{P}\right)=
- \frac{1}{W^0}\left(\vec{J}\times\vec{W}+\xi\vec{P}\right),
\end{equation}
where $\xi:=\vec{J}\cdot\vec{L}$ and $-W^0=\vec{J}\cdot\vec{P}$ is found in (\ref{41}).
From (\ref{19}) and (\ref{b10}) we obtain
\begin{equation}
 \label{b11}
 \frac{d}{dt}\left(W^0\xi\right)=-d(b-a)P^0\left(-\frac{1}{4(P^0)^2}(W^0)^4+\right.
 \end{equation}
 $$+
 \left(\frac{c}{d}+\frac{b-a}{2}\vec{J}^{\;2}+\frac{h_2-c_2}{2(P^0)^2(b-a)}\right)(W^0)^2+
 $$$$
 +\left.
 c_2\vec{J}^{\;2}-
 \left(\frac{h_2-c_2}{2P^0(b-a)}-\frac{b-a}{2}P^0\vec{J}^{\;2}\right)\right).
$$
Solution of (\ref{b11}) is given by
\begin{equation}
 \label{b12}
 \xi(t)=-\frac{d(b-a)P^0}{W^0}\int_{t_0}^{t}
 \left(-\frac{1}{4(P^0)^2}(W^0)^4(s)+
 \right.
 \end{equation}
 $$+
 \left(\frac{c}{d}+\frac{b-a}{2}\vec{J}^{\;2}+\frac{h_2-c_2}{2(P^0)^2(b-a)}\right)(W^0)^2(s)+
 $$
 $$+\left.
 c_2\vec{J}^{\;2}-
 \left(\frac{h_2-c_2}{2P^0(b-a)}-\frac{b-a}{2}P^0\vec{J}^{\;2}\right)\right)ds.
$$
Substituting (\ref{b12}) into (\ref{b10}) we obtain  $\vec{L}=\vec{L}(t)$.

\section{The case of relativistic particle with spin}

In this section we will present in details the physical interpretation of
Hamiltonian system (\ref{17}-\ref{19}) integrated in the previous section. We
will restrict ourselves to the case when the deformation parameter $a\in [-1,0]$,
i.e. when $E_a(1,3)$ is Poincare or Galilean group.

In our considerations we will use the twistor description of  the space--time $\mathbb{E}^{1,3}_a$, when $a=-1$. So, let us begin from some necessary facts concerning the twistor theory approach to the symplectic geometry of the relativistic particle phase spaces, for the details see \cite{4}.

Let us recall that the twistor space $\mathbb{T}$ is $\mathbb{C}^4$ equipped
with in the Hermitian form $\Phi$ of the signature $(++--)$. The Grassmannian
$G(2,\mathbb{T})=:\mathbb{M}$ of the two--dimensional subspaces
$z\subset\mathbb{T}$ of the twistor space $\mathbb{T}$ is  the
complexification $\overline{\mathbb{M}}^{\mathbb{C}(1,3)}$ of the conformal
compactification  $\overline{\mathbb{M}}^{1,3}$ of the Mincowski space
$\mathbb{M}^{1,3}$, which in our notation corresponds to
$\mathbb{E}_a^{1,3}$, with $a=-1$. One can enumerate the orbits
$\mathbb{M}^{k,l}$ of the action of the conformal group $SU(2,2)$ on
$\mathbb{M}$ by signatures $sign \Phi\mid_{z}=:(k,l)$ of the restrictions
$\Phi\mid_{z}$ of twistror forms $\Phi$ to subspace $z\in\mathbb{M}$. The orbit
$\mathbb{M}^{00}$ is identified with $\overline{\mathbb{M}}^{1,3}$.

In the following considerations we will take
\begin{equation}
\label{42k}
\Phi=i\left(\begin{array}{cc}
{\bf 0} & -\sigma_0\\
\sigma_0 & {\bf 0}
\end{array}\right),
\end{equation}
assume
\begin{align}
 \label{42}
 \infty:= & \left\{\left(\begin{array}{c}
                                        \eta\\
                                        0
                                       \end{array}\right)\in\mathbb{T}:\eta\in\mathbb{C}^2\right\}\in\mathbb{M}^{00},\\
 o:= & \left\{\left(\begin{array}{c}
                                        0\\
                                        \xi
                                       \end{array}\right)\in\mathbb{T}:\xi\in\mathbb{C}^2\right\}\in\mathbb{M}^{00},\nonumber
\end{align}
where ${\bf 0}=\left(\begin{array}{cc}
                0 & 0\\
                0 & 0
               \end{array}\right)$,
 $\sigma_0=\left(\begin{array}{cc}
                1 & 0\\
                0 & 1
               \end{array}\right)\in Mat_{2\times 2}(\mathbb{C})$, and let us identify
Lie algebra  $su(2,2)$ with its dual $su(2,2)^{*}$ by
\begin{equation}
 \label{b42}
 \left< X,\rho\right>:=Tr\left( X\rho\right), \;\;\;\;\rho, X\in su(2,2).
\end{equation}
The decomposition $\mathbb{T}= \infty\oplus o$ of the twistor space defines the corresponding decomposition of $su(2,2)\subset End\;\mathbb{T}$:
\begin{equation}
\label{b42a}
 su(2,2)\cong su(2,2)^*={\cal T}_{\infty}\oplus{\cal L}_{o,\infty}\oplus {\cal D}_{o,\infty}\oplus{\cal A}_o,
\end{equation}
where $X\in su(2,2)$ belongs to: the  translation Lie subalgebra ${\cal
T}_{\infty}\subset su(2,2)$ iff $Im X\subset\infty\subset Ker X$, the acceleration
Lie  subalgebra ${\cal A}_{\infty}$ iff  $Im X\subset o\subset Ker X$, while the
Lorentz subalgebra ${\cal L}_{o,\infty}$ and dilatation subalgebra ${\cal
D}_{o,\infty}$ are the commutant and centralizer of $su(2,2)_o\cap
su(2,2)_{\infty}$, respectively. Using the pairing (\ref{b42}) we obtain the
following vector space isomorphism:
\begin{equation}
 {\cal T}_{\infty}^*\cong {\cal A}_o,\;\;\;{\cal L}_{o,\infty}^*\cong {\cal L}_{o,\infty},\;\;\;{\cal D}_{o,\infty}^*\cong {\cal D}_{o,\infty},\; \textrm{ and }\; {\cal A}_o^*\cong {\cal T}_{\infty}.
\end{equation}
Now, let $(\mathbb{P},\omega)$ be $SU(2,2)$  -- symplectic manifold and let
${\cal J}:\mathbb{P}\longrightarrow su(2,2)^*\cong su(2,2)$ be the
corresponding momentum map of the phase space $\mathbb{P}$ into
Lie--Poisson space $su(2,2)^*$. Matrix block notation consistent with the
decomposition (\ref{b42a}) allows us to express ${\cal J}(p)$ at
$p\in\mathbb{M}$ as follows
\begin{equation}
\label{b42aa}
 {\cal J}(p)=
\left(
 \begin{array}{cc}
 \frac 12d(p)\sigma_0+M(p) & A(p)\\
 P(p) & -\frac 12 d(p)\sigma_0-M^{\dagger}(p)
                   \end{array}
                   \right),
\end{equation}
where $d(p)$ is the dilatation and
\begin{align}
\label{b42b}
 P(p)=P^{\mu}(p)\sigma_{\mu}, &&  A(p)=A^{\mu}(p)\sigma_{\mu},&& M(p)=\frac 12 M^{0k}\sigma_k+\frac i2 M^{kl}\epsilon_{kln}\sigma_n&,
 \end{align}
 where $L_k(p)=M^{0k}$, $ J_k(p)=\epsilon_{kln}M^{ln}$,
are the four--momentum, four--acceleration, and relativistic angular momentum,
respectively. In (\ref{b42b}) we use summation convention for $\mu=0,1,2,3$
and $k,l,n=1,2,3$ and $\sigma_1=\left(\begin{array}{cc}
0 & 1\\
1 & 0
\end{array}\right)$,
$\sigma_2=\left(\begin{array}{cc}
0 & i\\
-i & 0
\end{array}\right)$,
$\sigma_3=\left(\begin{array}{cc}
1 & 0\\
0 & -1
\end{array}\right)$ are the Pauli matrices. The Poincare Lie-Poisson space ${\cal P}^*(1,3)\cong{\cal T}_{\infty}^*\oplus{\cal L}^*_{o,\infty}$ is distinguished in $su(2,2)^*$ by $d=0$ and $A=0$. So, the momentum map ${\cal J}:p\longmapsto {\cal J}(p)\in{\cal P}^*(1,3)$ of the phase space $\mathbb{P}$ into ${\cal P}^*(1,3)$ is defined by
${\cal J}:p\longmapsto (P(p),M(p))\in Mat_{2\times 2}(\mathbb{C})\times
Mat_{2\times 2}(\mathbb{C})$. By ${\cal P}(1,3)$ we will denote Poincare--Lie
algebra which corresponds to $a=-1$.

For $(2\times 2)$--matrix calculus it is useful to introduce the following
operation on $B\in Mat_{2\times 2}(\mathbb{C})$:
\begin{equation}
\label{b42c}
\widetilde{B}:=\sigma_2 B^{\intercal}\sigma_2.
\end{equation}
For example using this operation we arrive at the identities
\begin{equation}
 \label{b42cc}
B\widetilde{B}= \det B\sigma_0\;\;\;\;\widetilde{AB}=\widetilde{B}\widetilde{A}
\end{equation}
and
\begin{equation}
 \det(A+B)=\det A+\det B+\textrm{Tr} \widetilde{A}B.
\end{equation}
We define the Pauli--Lubansky four--vector $W=W^{\mu}\sigma_{\mu}$ in the
following way
\begin{equation}
\label{b42d}
M\widetilde{P}=:R-iW,
\end{equation}
where $R^{\dagger}=R$ and $W^{\dagger}=W$.

Now we apply the formalism investigated in the above preliminaries to the
description of the Hamilton dynamics given in Section 2 in terms of the
Minkowski space geometry. Let us begine from the case of massless particle
with non-zero helicity. The phase space of such particle is the manifold the
positive defined projective twistors
\begin{equation}
 \label{b42e}
\mathbb{PT}^{+}:=\left\{[v]\in\mathbb{CP}(3):v^{\dagger}\Phi v>0\right\},
\end{equation}
where $[v]:=\mathbb{C}v$ is a one--dimensional complex subspace of
$\mathbb{T}$ spanded by $0\neq v\in\mathbb{T}$. The $SU(2,2)$-- invariant
symplectic form $\omega^{+}_{\alpha}$ on $\mathbb{PT}^{+}$ is the K\"ahler
form
\begin{equation}
 \label{b42f}
\omega^{+}_{\alpha}:=i\alpha\partial\overline{\partial}\log v^{\dagger}\Phi v.
\end{equation}

In the subsequent we will use spinor coordinates
$(\eta,\xi)\in\mathbb{C}^2\times\mathbb{C}^2$ for the twistor
$v=(\eta,\xi)\in\mathbb{T}$ defined by the decomposition
$\mathbb{T}=\infty\oplus o$. After passing to the homogeneous coordinates
$\zeta_1:=\dfrac{\eta_1}{\xi_2}$, $\zeta_2:=\dfrac{\eta_2}{\xi_2}$,
$\zeta:=\dfrac{\xi_1}{\xi_2}$, where $\xi_2\neq 0$, we obtain the coordinate
representation for symplectic form (\ref{b42f}):
\begin{equation}
 \label{46}
\omega^{+}_{\alpha}=\frac{-i\alpha}{\Delta^2}
\left(\begin{array}{ccc}
d\overline{\zeta}_1 & d\overline{\zeta}_2 & d\overline{\zeta}
        \end{array}\right)
\wedge\left(\begin{array}{ccc}
-\zeta\overline{\zeta} & -\zeta & \overline{\zeta}\zeta_1+\zeta_2-\overline{\zeta}_2\\
-\overline{\zeta} & -1 & \overline{\zeta}_1\\
\zeta\overline{\zeta}_1-\zeta_2+\overline{\zeta}_2 & \zeta_1 & -\zeta_1\overline{\zeta}_1
\end{array}\right)
\left(\begin{array}{c}
         d\zeta_1\\
         d\zeta_2\\
         d\zeta
        \end{array}\right),
\end{equation}
where
\begin{equation}
 \label{45}
\Delta=v^{\dagger}\Phi v=i\left(\overline{\zeta}\zeta_1-\zeta\overline{\zeta}_1+\zeta_2-\overline{\zeta}_2\right).
\end{equation}
The momentum map ${\cal
J}^{+}_{\alpha}:(\mathbb{PT}^{+},\omega^{+}_{\alpha})\longrightarrow
su(2,2)^{*}$ for the symplectic manifold
$(\mathbb{PT}^{+},\omega^{+}_{\alpha})$ is the following one
\begin{equation}
 \label{b45}
{\cal J}^{+}_{\alpha}([v])=i\alpha\left(\frac 14 {\bf 1}-\dfrac{vv^{\dagger}\Phi}{\Delta}\right),
\end{equation}
and it leads to the formulas for four-momentum and relativistic angular
momentum
\begin{align}
\label{b45a} & P([v])=\frac{i\alpha}{{\eta}^{\dagger}{\xi}-{\xi}^{\dagger}{\eta}}{\xi}{\xi}^{\dagger},\\
\label{b45b} & M([v])=\frac{i\alpha}{{\eta}^{\dagger}{\xi}-{\xi}^{\dagger}{\eta}}
\left({\eta}{\xi}^{\dagger}-\frac 12{\xi}^{\dagger}{\eta}\sigma_0\right).
\end{align}
Using (\ref{b45a}) and (\ref{b45b}) we obtain $P^{\mu}([v])$, $\vec{L}([v])$,
$\vec{J}([v])$ and $W^{\mu}([v])$ in the $(\zeta_1, \zeta_2,\zeta)$ --coordinate
representation:
\begin{align}
\label{49} & P^0=  -\frac{i\alpha}{2\Delta}(\zeta\overline{\zeta}+1),\\
\label{50} & P^1=  -\frac{i\alpha}{2\Delta}(\zeta+\overline{\zeta}),\\
\label{51} & P^2=  -\frac{\alpha}{2\Delta}(\overline{\zeta}-\zeta),\\
\label{52} & P^3=  -\frac{i\alpha}{2\Delta}(\zeta\overline{\zeta}-1),
\end{align}
\begin{align}
\label{53} & L_1=  \frac{\alpha}{2\Delta}(\zeta_2\overline{\zeta}+\overline{\zeta}_2\zeta+\zeta_1+\overline{\zeta}_1),\\
\label{54} & L_2=  \frac{i\alpha}{2\Delta}(-\zeta_2\overline{\zeta}+\overline{\zeta}_2\zeta+\zeta_1-\overline{\zeta}_1),\\
\label{55} & L_3=  \frac{\alpha}{2\Delta}(\zeta_1\overline{\zeta}+\overline{\zeta}_1\zeta-\zeta_2-\overline{\zeta}_2),\\
\label{56} & J_1=  -\frac{i\alpha}{2\Delta}(\zeta_2\overline{\zeta}-\overline{\zeta}_2\zeta+\zeta_1-\overline{\zeta}_1),\\
\label{57} & J_2=  -\frac{\alpha}{2\Delta}(\zeta_2\overline{\zeta}+\overline{\zeta}_2\zeta-\zeta_1-\overline{\zeta}_1),\\
\label{58} & J_3=  -\frac{i\alpha}{2\Delta}(\zeta_1\overline{\zeta}-\overline{\zeta}_1\zeta-\zeta_2+\overline{\zeta}_2),\\
\label{62} & W^0=  \frac{i\alpha^2}{4\Delta}(\zeta\overline{\zeta}+1),\\
\label{63} & W^1=  -\frac{i\alpha^2}{4\Delta}(\zeta+\overline{\zeta}),\\
\label{64} & W^2=  -\frac{\alpha^2}{4\Delta}(\overline{\zeta}-\zeta),\\
\label{65} & W^3=  -\frac{i\alpha^2}{4\Delta}(\zeta\overline{\zeta}-1).
\end{align}
The momentum map (\ref{b45}) is  a Poisson map from the symplectic manifold
$(\mathbb{PT}^{+},\omega^{+}_{\alpha})$ into Lie--Poisson space $({\cal
P}(1,3)^{*},\{\cdot ,\cdot \}_{a})$, i.e. for $f,g\in C^{\infty}({\cal P}(1,3)^{*})$ we
have
\begin{equation}
\label{59}
\{f,g\}_{a}\circ {\cal J}^{+}=\{ f\circ{\cal J}^{+},g\circ{\cal J}^{+} \}_{\alpha,+},
\end{equation}
where Poisson bracket $\{\cdot ,\cdot \}_{\alpha,+}$ is defined by the
symplectic form $\omega^{+}_{\alpha}$. In the coordinates $(\zeta_1,
\zeta_2,\zeta)$ it takes the form
\begin{align}
 \label{47}
\{F,G\}_{\alpha,+}= & -\frac{\Delta}{\alpha}\left(\zeta_1
\left(\frac{\partial F}{\partial\overline{\zeta}_2}\frac{\partial G}{\partial \zeta_1}-\frac{\partial F}{\partial\zeta_1}\frac{\partial G}{\partial \overline{\zeta}_2}\right)
-\overline{\zeta}_1
\left(\frac{\partial F}{\partial\overline{\zeta}_1}\frac{\partial G}{\partial \zeta_2}-\frac{\partial F}{\partial\zeta_2}\frac{\partial G}{\partial \overline{\zeta}_1}\right)\right. +\nonumber\\
& +\left.\left(\zeta_2-\overline{\zeta}_2\right)
\left(\frac{\partial F}{\partial\overline{\zeta}_2}\frac{\partial G}{\partial \zeta_2}-\frac{\partial F}{\partial\zeta_2}\frac{\partial G}{\partial \overline{\zeta}_2}\right)+
\frac{\partial F}{\partial\overline{\zeta}}\frac{\partial G}{\partial \zeta_1}-\frac{\partial F}{\partial\zeta_1}\frac{\partial G}{\partial \overline{\zeta}}+\right.\nonumber\\
& +\left.\frac{\partial F}{\partial\zeta}\frac{\partial G}{\partial \overline{\zeta}_1}-\frac{\partial F}{\partial\overline{\zeta}_1}\frac{\partial G}{\partial \zeta}+
\overline{\zeta}
\left(\frac{\partial F}{\partial\zeta_2}\frac{\partial G}{\partial \overline{\zeta}}-\frac{\partial F}{\partial\overline{\zeta}}\frac{\partial G}{\partial \zeta_2}\right)\right.+\\
& \left. +\zeta
\left(\frac{\partial F}{\partial\overline{\zeta}_2}\frac{\partial G}{\partial \zeta}-\frac{\partial F}{\partial\zeta}\frac{\partial G}{\partial \overline{\zeta}_2}\right)
\right),\nonumber
\end{align}
for $F,G\in C^{\infty}(\mathbb{PT}^{+})$. The four-momentum $P^{\mu}([v])$
and Pauli--Lubansky vector $W^{\mu}([v])$ defined in (\ref{b45a}), (\ref{b45b})
and (\ref{b42d}) satisfies the relationships
\begin{equation}
\label{b47r}
\eta^a_{\mu\nu}P^{\mu}([v]) P^{\nu}([v])=0\;\;\;\;\textrm{and}\;\;\;\; W^{\mu}([v])=\frac{\alpha}{2} P^{\mu}([v]).
\end{equation}

The above conditions confirm that we are in the case of the massless particle
with the helicity $\dfrac{\alpha}{2}$ which phase space is given by the
symplectic leaves of the Lie--Poisson bracket $\{\cdot ,\cdot\}_{a}$ defined by
the conditions $c_1=c_2=0$. So, one can pull back the solution
$\left(\vec{P}(t),\vec{L}(t)\right)$ of Hamilton equations (\ref{17}-\ref{19}) on the
symplectic manifold $\left(\mathbb{PT}^{+},\omega^{+}_{\alpha}\right)$ by the
map
\begin{align}
\label {b47b} & \zeta_1=\dfrac{L_2-J_1+i(L_1+J_2)}{2(P^3-P^0)},\nonumber\\
 & \zeta_2=\dfrac{(P^1+iP^2)(L_2-J_1+i(L_1-J_2))-(P^3-P^0)(J_3-iL_3)}{-2(P^3-P^0)^2},\\
 & \zeta=\dfrac{P^1-iP^2}{P^0-P^3}.\nonumber
\end{align}
So, if $P^0(t)$, $\vec{P}(t)$, $\vec{L}(t)$, $\vec{J}(t)$ satisfy equations
(\ref{17}-\ref{19}) then $\zeta_1(t),\zeta_2(t),\zeta(t))$, given by (\ref{b47b}), are
the solution of  Hamilton equations
\begin{align}
\label{b47c}& \frac{d}{dt}\zeta_1(t)=\{\zeta_1, h\circ {\cal J}^{+}_{\alpha} \}_{\alpha,+}=
-\frac{\alpha}{4\Delta}(b-a)(\zeta\bar{\zeta}+1)\left(\left(c-\frac{d\alpha^2}{4}\right)\zeta-\right.\\
& -\frac{\alpha^2}{4\Delta^2}d(b-a)
\left(-6\zeta_2\bar{\zeta}_2\zeta^2\bar{\zeta}
+4\zeta_1\zeta_2\zeta\bar{\zeta}
-4\zeta_1\bar{\zeta}_1\zeta^2\bar{\zeta}
+4\zeta_1\bar{\zeta}_2\zeta\bar{\zeta}+\nonumber
\right.\\
& +3\zeta^2_1\zeta\bar{\zeta}^2
+2\bar{\zeta}_1\bar{\zeta}_2\zeta^2
+2\bar{\zeta}_1\zeta_2\zeta^2
-4\zeta_2\bar{\zeta}_2\zeta
+\bar{\zeta}_1^2\zeta^3+\zeta_2^2\zeta+\bar{\zeta}_2^2\zeta+\nonumber\\
& \left.\left.+2\zeta_1^2\bar{\zeta}
+2\zeta_1\bar{\zeta}_2
+2\zeta_1\zeta_2\zeta^2\bar{\zeta}^2
+2\zeta_1\bar{\zeta}_1\zeta^2\bar{\zeta}\right)\right)\nonumber
,\\
& \frac{d}{dt}\zeta_2(t)=\{\zeta_2, h\circ {\cal J}^{+}_{\alpha} \}_{\alpha,+}=
\frac{i}{2}(b-a)P^0\left(c-d\frac{\alpha^2}{4}+d(b-a)\vec{J}^2\right)\left(\zeta\bar{\zeta}-1\right)+\nonumber\\
&
+\frac{\alpha d}{4\Delta}(b-a)^2(P^0)^2
\left(
-2\zeta_1\bar{\zeta_1}\zeta\bar{\zeta}-2\zeta_1\bar{\zeta}_1-4\zeta_2\bar{\zeta}_2\zeta\bar{\zeta}-2\zeta_2\bar{\zeta}_2+2\zeta_1\bar{\zeta}_2\bar{\zeta}-\bar{\zeta}_2+\right.\\
&+\zeta_2^2+\bar{\zeta}_2^2+2\zeta_2^2\zeta\bar{\zeta}+2\bar{\zeta}_1\bar{\zeta}_2\zeta+\bar{\zeta}_1^2\zeta^2+\zeta_1^2\bar{\zeta}^2+2\bar{\zeta}_2\zeta-2\zeta_1-\zeta_1\bar{\zeta}+\bar{\zeta}_1\zeta-\zeta_2
\left. \right)
,\nonumber\\
\label{b47d}& \frac{d}{dt}\zeta(t)=\{\zeta, h\circ {\cal J}^{+}_{\alpha} \}_{\alpha,+}=-\frac{\alpha^3}{8\Delta^3}d(b-a)^2(\zeta\bar{\zeta}+1)^3(\zeta_2\zeta-\zeta_1),
\end{align}
defined on $\left(\mathbb{PT}^{+},\omega^{+}_{\alpha}\right)$ by the
Hamiltonian $h\circ {\cal J}^{+}_{\alpha}$, where $h$ is given by (\ref{15}).
Equations (\ref{b47c}-\ref{b47d}) prove to form a system with rational
nonlinearities. Due to (\ref{b47r}), it follows from the above considerations that
after passing to the coordinates $(\vec{P},\vec{W})$  we linearize the
hamiltonian system under consideration.

In order to describe the phase space of a massive particle with the spin $s\neq 0$ let us consider twistor flag space
\begin{equation}
 \label{48}
 \mathbb{F}:=\left\{([v],z)\in\mathbb{PT}\times \mathbb{M}:[v]\subset z\right\}.
\end{equation}
Similarly to the case of Grassmannian $\mathbb{M}$ we will enumerate the
orbits $\mathbb{F}^{k,lm}$ of the natural action of $SU(2,2)$ on $\mathbb{F}$
by the signatures $k=sign\; \Phi\mid_{[v]}$, $lm=sign\;\Phi\mid_{z}$ of the
restrictions of twistor form to the flag $[v]\subset z$. We restrict our interest to
the  orbit $\mathbb{F}^{+,++}$ consisting of the positive flags. For physical
interpretation of the flag spaces of differen signatures, see \cite{4}. One has
the following double fibration

\hspace*{2cm}
\xymatrix{
 & & \mathbb{F}^{+,++} \ar@<.0ex>[ddrr]^*-<.9ex>\txt{\tiny{$\pi_{2}$}}
             \ar@<-.0ex>[ddll]^*-<4ex>\txt{\tiny{$\pi_{1}$}}      & &
             &  \\
             & &&&    &\\
\mathbb{PT}^{+}     &&&     & \mathbb{M}^{++}
 }
 \\
of $\mathbb{F}^{+,++}$  over $\mathbb{PT}^{+}$ and $\mathbb{M}^{++}$.
According to \cite{4} we show now that $\mathbb{M}^{++}$ is the phase space
of massive spinless particle and $\mathbb{F}^{+,++}$ is the phase space of a
massive particle with non-zero spin. For these reasons let us pass to the
coordinate description of $\mathbb{M}^{++}$ and $\mathbb{F}^{+,++}$
consistent with the decomposition $\mathbb{T}=\infty\oplus o$ given by
(\ref{42}). We have
\begin{equation}
 \label{49k}
 v=\left(\begin{array}{c}
          Z\xi\\
          \xi
         \end{array}\right)\;\;\;\;\textrm{and}\;\;\;\;z=\left\{\left(\begin{array}{c}
         Z\xi\\
         \xi
         \end{array}\right):\;\xi\in\mathbb{C}^2\right\}
\end{equation}
for $[v]\subset z$, where $Z\in Mat_{2\times 2}(\mathbb{C})$. The flag $[v]\subset z$ belongs to $\mathbb{F}^{+,++}$ iff
the "imaginary" part of $Z=X+iY$, $X^{\dagger}=X$ and $Y^{\dagger}=Y$, is positive definite, i.e.
\begin{equation}
 \label{50k}
 \det Y>0 \;\;\;\;\textrm{and}\;\;\;\;\textrm{Tr} \;Y>0.
\end{equation}

Let us take the product $\mathbb{PT}^{+}\times\mathbb{PT}^{+}$ of the one--twistor phase spaces with the symplectic form
\begin{equation}
 \label{51k}
\omega^{12}=\pi_1^*\omega_{\alpha_1}^{+}\oplus\pi_2^*\omega_{\alpha_2}^{+},
\end{equation}
where $\pi_i:\mathbb{PT}^{+}\times\mathbb{PT}^{+}\longrightarrow
\mathbb{PT}^{+}$ is the projection on the i--th component of the product. The
symplectic form $\omega^{12}$ is invariant with respect to the natural action of
$SU(2,2)$ on $\mathbb{PT}^{+}\times\mathbb{PT}^{+}$ and the momentum
map ${\cal J}^{1,2}:\mathbb{PT}^{+}\times\mathbb{PT}^{+}\longrightarrow
su(2,2)^{*}$ for $(\mathbb{PT}^{+}\times\mathbb{PT}^{+},\omega^{12})$ is
given by
\begin{equation}
\label{52k}
{\cal J}^{1,2}={\cal J}^{+}_{\alpha}\circ\pi_1+{\cal J}^{+}_{\alpha}\circ\pi_2.
\end{equation}
The function $s^2:\mathbb{PT}\times\mathbb{PT}\longrightarrow\mathbb{R}$ defined by
\begin{equation}
 \label{53k}
s^2([v_1],[v_2]):=\left(\frac{\alpha_1-\alpha_2}{4}\right)^2+\frac{\alpha_1\alpha_2}{4}\frac{|v_1^{\dagger}\Phi v_2|^2}{v_1^{\dagger}\Phi v_1v_2^{\dagger}\Phi v_2}
\end{equation}
in an invariant of the conformal group $SU(2,2)$ . The projective twistors are orthogonal $[v_1]\bot [v_2]$
with respect to the twistor form $\Phi$ iff
\begin{equation}
 \label{54k}
s^2([v_1],[v_2]):=\left(\frac{\alpha_1-\alpha_2}{4}\right)^2.
\end{equation}
Any flag $[v]\subset z$ one can identify with the pair of twistors $([v_1],[v_2])\in \mathbb{PT}^{+}\times\mathbb{PT}^{+}$ satisfying condition (\ref{54k}). Namely, one puts $z=span\{[v_1],[v_2]\}\in\mathbb{M}$ and $[v]=[v_1]$. Reducing symplectic form $\omega^{1,2}$ to the level submanifold of $\mathbb{PT}^{+}\times\mathbb{PT}^{+}$ defined by (\ref{54k}) we obtain $SU(2,2)$ -- invariant symplectic form (K\"ahler form) $\omega_{s,\delta}$ on $\mathbb{F}^{+,++}$ which in the coordinates $([\xi],Z)\in\mathbb{CP}(1)\times Mat_{2\times 2}(\mathbb{C})$ is given by
\begin{equation}
 \label{55k}
\omega_{s,\delta}=i\partial \overline{\partial}
\log\left[\left(\det\left(Z-Z^{\dagger}\right)\right)^{s+2\delta}\left(\eta^{\dagger}\left(Z-Z^{\dagger}\right)\eta\right)^{4s}\right],
\end{equation}
where $s:=\frac{\alpha_1-\alpha_2}{4}$ and $\delta:=-\frac{\alpha_1+\alpha_2}{4}$.
The symplectic form (\ref{55k}) rewrited in the variables $P^{\mu}$, $W^{\mu}$, $X^{\mu}$ is the Souriau symplectic form, see \cite{11}.

The reduced momentum map ${\cal
J}_{s,\delta}:\mathbb{F}^{+,++}\longrightarrow su(2,2)^{*}$ is of the form
\begin{equation}
 \label{56k}
{\cal J}_{s,\delta}([\xi],Z)=\left(\begin{array}{cc}
ZP-i\delta\sigma_0 & -ZPZ^{\dagger}\\
P & -PZ^{\dagger}-i\delta\sigma_0
                                   \end{array}\right),
\end{equation}
where
\begin{equation}
 \label{57k}
P=-\frac{i\alpha_1}{\xi^{\dagger}(Z-Z^{\dagger})\xi}\xi\xi^{\dagger}
-\frac{i\alpha_2}{det(Z-Z^{\dagger})\xi^{\dagger}(Z-Z^{\dagger})\xi}
(\widetilde{Z}-\widetilde{Z}^{\dagger})\widetilde{\xi\xi^{\dagger}}(\widetilde{Z}-\widetilde{Z}^{\dagger}).
\end{equation}
Comparing  (\ref{b42aa}) with (\ref{56k}) we obtain
\begin{equation}
 \label{58k}
M=ZP-\frac 12\textrm{Tr}(ZP)\sigma_0.
\end{equation}
Substituting (\ref{57k}) into (\ref{58k}) we also express $M$ in the coordinates
$([\xi],Z=X+iY)$:
\begin{equation}
 \label{58kk}
M=
-\frac{i\alpha_1}{\xi^{\dagger}(Z-Z^{\dagger})\xi}Z\xi\xi^{\dagger}
+\frac{i\alpha_1}{2\xi^{\dagger}(Z-Z^{\dagger})\xi}\xi^{\dagger}Z\xi\sigma_0-
\end{equation}
$$-\frac{i\alpha_2}{det(Z-Z^{\dagger})\xi^{\dagger}(Z-Z^{\dagger})\xi}
(\widetilde{Z}-\widetilde{Z}^{\dagger})\widetilde{\xi\xi^{\dagger}}(\widetilde{Z}-\widetilde{Z}^{\dagger})+
$$
$$
+\frac{i\alpha_2}{2det(Z-Z^{\dagger})\xi^{\dagger}(Z-Z^{\dagger})\xi}
\overline{
\xi^{\dagger}(Z-Z^{\dagger})\sigma_2\overline{Z}\sigma_2(Z-Z^{\dagger})
\xi
}\sigma_0.
$$

From equalities  (\ref{57k}-\ref{58k}), definition (\ref{b42d}) and identities
(\ref{b42cc}) we find that
\begin{equation}
 \label{59k}
W=
-\frac{i\delta\alpha_1}{\xi^{\dagger}(Z-Z^{\dagger})\xi}
\widetilde{\xi\xi^{\dagger}}
-\frac{i\alpha_1\alpha_2}{2det(Z-Z^{\dagger})}(Z-Z^{\dagger})-
\end{equation}
$$
-\frac{i\delta\alpha_2}{det(Z-Z^{\dagger})\xi^{\dagger}(Z-Z^{\dagger})\xi}
(Z-Z^{\dagger})\xi\xi^{\dagger}(Z-Z^{\dagger}).
$$
and
\begin{align}
 \label{60k} & \textrm{Tr}\;PW=0,\nonumber\\
 & \det W=-s^2\det P,\\
& \textrm{Tr }PY=2\delta.\nonumber
\end{align}
Using vector notation for $M$, see (\ref{b42b}), we can rewrite  (\ref{58k}) as follows
\begin{align}
 \label{62k} & \vec{L}=X_0\vec{P}+P^0\vec{X}-\vec{Y}\times\vec{P},\\
 & \vec{J}=Y_0\vec{P}+P^0\vec{Y}+\vec{X}\times\vec{P}.\nonumber
\end{align}
Inverting (\ref{58k}) we find
\begin{align}
 \label{64k} & Y_0=-\frac{1}{\det P}\left(W^0-\delta P^0\right),\\
\label{65k} & \vec{Y}=-\frac{1}{\det P}\left(\vec{W}+\delta \vec{P}\right),\\
 \label{66k} & \vec{X}=\frac{1}{\det P}\vec{J}\times\vec{P}+
\frac{P_0}{\det P}\vec{L}-
\frac{1}{P^0\det P}\left(\left(\vec{P}\cdot\vec{L}\right)+\det PX_0\right)\vec{P}.
\end{align}

The formula given above allows us  to obtain the time evolution $Y_0=Y_0(t)$,
$\vec{Y}=\vec{Y}(t)$ and $\vec{X}=\vec{X}(t)$  described by the Hamiltonian
(\ref{16}). For this reason we only need to assume that the evolution parameter
$t$ appearing in the Hamilton equations (\ref{17}-\ref{19}) is the time related to
the space--time coordinate $X_0$ by $X_0=ct$, where $c$ is the light velocity.
From (\ref{66k}) we have
\begin{equation}
 \label{67k}
  \vec{X}=-\frac{1}{(mc)^2}\vec{J}\times
  \left(\vec{P}(t)-\frac{P^0}{W^0(t)}\vec{W}(t)\right)
  +
\end{equation}
$$
+\left(ct+\frac{1}{(a-b)dP^0}\frac{d}{dt}\ln W^0(t)+(mc)^2 \xi(t) \right)\frac{\vec{P}(t)}{P^0},
$$
where $m$ is  the relativistic particle mass defined by $-(mc)^2=c_1$, $cP^0$
and $\vec{J}$ are its energy and angular momentum, being integral of motions
in the case under consideration. Note that $\xi(t)$ is expressed in terms of
$W^0(t)$ in (\ref{b12}) while  elliptic function $W^0(t)$  is defined in (\ref{41}).
Time evolution of the momentum $\vec{P}(t)$ and spin $\vec{W}(t)$ is also
given in terms of elliptic function $W^0(t)$, see formula (\ref{b6}-\ref{b9}).

The physical sense of the integrated Hamiltonian dynamics one can recognize
from the form of the total energy function (\ref{16}).

\section{Concluding remarks}

The hamiltonian integrable system investigated in the paper describes time
evolution of the relativistic particle (massive and massless) with non--zero
spin. The Hamiltonian governing this evolution is invariant with respect to the
translation of  Minkowski space-time and the rotation of the space, which
means that the particle does not interact with the external filed. However, the
dynamics has rather complicated nonlinear character that is caused by the
coupling between the momentum,  angular momentum and spin of the  particle.

The Galilean case  we do not discuss separately. It can be obtained from
relativistic one in the $a$--vanishing limit.

The Euclidean case needs  an approach different from the twistor one. Here
we can consider four-vectors $W^{\mu}$ and $P^{\mu}$ as a dynamical
variables which satisfy the conditions (\ref{12}), (\ref{13}) and (\ref{b4cc}). As a
result one obtains a hamiltonian system on the bundle $T\mathbb{S}^3$ of
vector spaces tangent to  three--dimensional sphere $\mathbb{S}^3$. In the
paper we do not  discuss this case in details.


\bibliographystyle{plain}

\end{document}